\documentclass[preprint,showpacs,preprintnumbers,amsmath,amssymb]{revtex4}
\usepackage{graphicx}
\usepackage{dcolumn}
\usepackage{bm}
\usepackage{mathrsfs}
\usepackage{amsmath,amssymb,graphics}

\begin{document}

 \title{Anomaly analysis of Hawking radiation from
  Kaluza-Klein black hole with squashed horizon}

\author{Shao-Wen Wei\footnote{E-mail: weishaow06@lzu.cn} ,
        Ran Li\footnote{E-mail: liran05@lzu.cn},
        Yu-Xiao Liu\footnote{Corresponding author. E-mail: liuyx@lzu.edu.cn}
        and Ji-Rong Ren}
\affiliation{
    Institute of Theoretical Physics, Lanzhou University,
           Lanzhou 730000, P. R. China}

\begin{abstract}
Considering gravitational and gauge anomalies at the horizon, a
new method that to derive Hawking radiations from black holes has
been developed by Wilczek et al. In this paper, we apply this
method to non-rotating and rotating Kaluza-Klein black holes with
squashed horizon, respectively. For the rotating case, we found
that, after the dimensional reduction, an effective $U(1)$ gauge
field is generated by an angular isometry. The results show that
the gauge current and energy-momentum tensor fluxes are exactly
equivalent to Hawking radiation from the event horizon.
 \end{abstract}

\pacs{ 04.62.+v, 04.70.Dy, 11.30.-j \\
  Keywords: Hawking radiation, gauge and gravitational anomalies}

 \maketitle

 \section{introduction}

Hawking \cite{hawking} has been derived that black hole can
radiate from the event horizon like a black body at the
temperature $T=\frac{\kappa}{2\pi}$ using the method of quantum
field theory in curved spacetime. This rises the interest of
investigation for the Hawking radiation. An elegant derivation is
the tunneling method, which based on pair creations of particles
and antiparticles near the horizon through calculating WKB
amplitudes for classically forbidden trajectories
\cite{Parikh:1999mf,tunneling}. It is generally believed that the
Hawking radiation should be determined only by some universal
quantum effects just on the horizon.

Recent proposal of deriving Hawking radiation via gravitational and
gauge anomalies proposed by Wilczek and his collaborators
\cite{Robinson2005prl,Iso2006prl} has been attracted a lot of
interests. This rejuvenates the interest of investigation for
Hawking radiation. For various types of black holes, these
investigations have been carried out
\cite{Vagenas2006jhep,Jiang2007plb,Murata2007prd,Miyamoto2008prd,
Ma0709.3684,Wushuangqing2007cqg,Pen0709.0044,
Wushuangqing0709.4074,Pen0709.0157,Gangopadhyay2008prd,Kulkarni2008cqg,Wushuangqing2008cqg,
Kim2008,Bonora2008jhep,Shirasaka2008}. In fact, the anomaly analysis
can be traced back to Christensen and Fulling's early work
\cite{Christensen1977prd}, in which they suggested that there exists
a relation between the Hawking radiation and anomalous trace of the
field under the condition that the covariant conservation law is
valid. Wilczek et al. showed that the Hawking fluxes of the black
hole can be accounted by the gauge and gravitational anomalies at
the horizon. Their basic idea is that, near the horizon, a quantum
field in a black hole background can be effectively described by an
infinite collection of (1+1)-dimensional fields on $(t, r)$ space,
where $r$ is the radial direction.
In this two-dimensional reduction, because all the ingoing modes can
not classically affect physics outside the horizon, the
two-dimensional effective action in the exterior region becomes
anomalous with respect to gauge or general coordinate symmetries. To
cancel the anomaly, they found the Hawking flux is universally
determined only by the value of anomalies at the horizon. The
anomaly method is believed to be universal and independent of the
gravity theory and the dimension of spacetime. On the other hand,
Banerjee and Kulkarni \cite{Banerjee2008prd,Banerjee2008plb} pointed
that the Hawking radiation can be obtained only use the covariant
expression.

In Ref. \cite{Wushuangqing2007cqg}, the original work of
\cite{Robinson2005prl,Iso2006prl} was extended to the more general
case where the metric determinant $\sqrt{-g}\neq 1$, and the case in
which $\sqrt{-g}$ vanishes at the horizon had also been investigated
\cite{Pen0709.0044} in details. Recently, in
\cite{Papantonopoulos2008}, a non-spherical topological black hole
was considered and the Hawking radiation via gravitational anomalies
was derived. The aim of this paper is to generalize the anomaly
method to non-rotating and rotating Kaluza-Klein black holes with
squashed horizon. At first, we will perform a dimensional reduction
of an action given by a scalar field minimally coupled to gravity in
the background of a (4+1)-dimensional Kaluza-Klein black hole. The
result shows that the theory is reduced to an effective theory of an
infinite collection of (1+1)-dimensional scalar fields near the
horizon. Then, through studying the gauge and gravitational
anomalies, we obtain the Hawking flux. Especially, for the rotating
case, we found that, after the dimensional reduction, an effective
$U(1)$ gauge field is generated by an angular isometry. The
azimuthal quantum number $\lambda$ serves as the charge of each
partial wave, and this result accords with that of
\cite{Isop2006prd}.

The paper is organized as follows. In section \ref{nonKK}, we review
the basic properties of the (4+1)-dimensional non-rotating
Kaluza-Klein black hole with squashed horizon and carry out the
dimensional reduction. It's Hawking radiation is derived via
anomalies. The same calculation is applied to the rotating
Kaluza-Klein black hole with squashed horizon in section
\ref{totKK}. Finally, the paper ends with a brief summary.

\section{Quantum field in non-rotating Kaluza-Klein black hole with squashed horizon}
\label{nonKK}

We firstly review some main results of the Kaluza-Klein black hole
with squashed horizon. Some relevant discussions can be found in
\cite{Ishihara2006ptp,Wang2006npb,Chen2008prd,Ishihara2007prd}.
The metric of Kaluza-Klein black hole with squashed horizon is
given by
\begin{eqnarray}
 ds^2=-f(r)dt^2+\frac{k^2(r)}{f(r)}dr^2+\frac{r^2}{4}k(r)d\Omega^2+\frac{r^2}{4}(d\psi+\textrm{cos}\theta d\phi)^2,
\end{eqnarray}
where $d\Omega^2=d\theta^2+\textrm{sin}^2\theta d\phi^2$ is the line
element on the unit sphere. The coordinate ranges are
\begin{eqnarray}
\theta\in[\;0, \pi)\;,\phi\in[\;0, 2\pi)\;,\psi\in[\;0, 4\pi).
\end{eqnarray}
The functions $f(r)$ and $k(r)$ take the following forms:
\begin{eqnarray}
 f(r)&=&\frac{(r^2-r_+^2)(r^2-r_-^2)}{r^4},\nonumber\\
 k(r)&=&\frac{(r_{\infty}^2-r_+^2)(r_{\infty}^2-r_-^2)}{(r_{\infty}^2-r^2)^2}.
\end{eqnarray}
There are three coordinate singularities $r_+$, $r_-$ and
$r_{\infty}$, which satisfy $r_- < r_+ < r_{\infty}$,
$0<r<r_{\infty}$. $r=r_+$ and $r=r_-$ are outer and inner event
horizons of the black hole. $r=r_\infty$ is the spatial infinity.
They relate to the mass $M$, charge $Q$ by
$M=\frac{3\pi(r_{+}^{2}r_{\infty}^{2}+r_{-}^{2}r_{\infty}^{2}-2r_{+}^{2}r_{-}^{2})}{8G\sqrt{(r_{\infty}^{2}-r_{+}^{2})(r_{\infty}^{2}-r_{-}^{2})}}$,
$Q=\frac{\sqrt{3}\pi r_{+}r_{-}}{2 G}$. It has been shown that the
shape of the horizon is a deformed sphere determined by the factor
$k(r_+)$. The non-vanishing gauge potential is given by
\begin{eqnarray}
 A_t=\pm\frac{\sqrt{3}}{2}r_+r_-\bigg(\frac{1}{r^2}-\frac{1}{r_{\infty}^2}\bigg).
\end{eqnarray}
In this paper, we use the metric
\begin{eqnarray}
 ds^2=-F(\rho)d\tau^{2}+\frac{K^2(\rho)}{F(\rho)}d\rho^2+\rho^2K^2(\rho)d\Omega^2
        +\frac{r_{\infty}^2}{4K^2(\rho)}(d\psi+\textrm{cos}\theta
 d\phi)^2,
\end{eqnarray}
which can be deduced from (1) via coordinate transformation
\begin{eqnarray}
 \tau=\frac{2\rho_0 t}{r_{\infty}},\;\;
 \rho=\frac{\rho_0 r^2}{(r_{\infty}^2-r^2)},
\end{eqnarray}
where
$\rho_{0}^{2}={(r_{\infty}^{2}-r_{+}^{2})(r_{\infty}^{2}-r_{-}^{2})}
/({4r_{\infty}^{2}})$. The new functions $F(\rho)$ and $K(\rho)$ are
given by
\begin{eqnarray}
 F=\bigg(1-\frac{\rho_+}{\rho}\bigg)\bigg(1-\frac{\rho_-}{\rho}\bigg),\;\;
 K^2=1+\frac{\rho_0}{\rho},
\end{eqnarray}
and $\rho_\pm=\rho_{0}r^{2}_{\pm}/(r^{2}_{\infty}-r^{2}_{\pm})$. In
this coordinate, the non-vanishing gauge potential $A_\tau$, the
black hole mass $M$, the charge $Q$ and the surface gravity $\kappa$
of the outer horizon are respectively given by
\begin{eqnarray}
 A_\tau&=&\pm\frac{\sqrt{3}}{2}\frac{\sqrt{\rho_+\rho_-}}{\rho},\nonumber\\
 M&=&\frac{3\pi r_{\infty}}{4G}(\rho_++\rho_-),\nonumber\\
 Q&=&\mp \frac{\sqrt{3}\pi
 r_\infty}{G}\sqrt{\rho_+\rho_-},\nonumber\\
 \kappa&=&\frac{(\rho_+-\rho_-)}{2\rho_+^2}\sqrt{\frac{\rho_+}{\rho_++\rho_0}}.\label{gravity}
\end{eqnarray}

Next, we consider a complex scalar field in the squashed
Kaluza-Klein black hole background. The action functional is given
by
\begin{eqnarray}
 S[\varphi]&=&-\frac{1}{2}\int d^5x \sqrt{-g} g^{\mu\nu}
  (D_\mu\varphi)^{*} D_\nu\varphi\nonumber\\
 &=&\frac{1}{2}\int d^5x \varphi^{*} (\partial_\mu-ieA_\mu)
 [\sqrt{-g}g^{\mu\nu}(\partial_\nu-ieA_\nu)\varphi]\nonumber\\
 &=&\frac{1}{2}\int d^5x \rho^2 K^2 r_{\infty}\sin\theta \;
 \varphi^{*} \bigg\{-\frac{1}{F}(\partial_\tau -ieA_\tau)^2\varphi
 +\frac{1}{\rho^2K^2}\partial_\rho\rho^2
 F\partial_\rho\varphi\nonumber\\
 &&+\frac{1}{\rho^2K^2}\bigg[
  \frac{1}{\textrm{sin}\theta}\partial_\theta(\textrm{sin}\theta\partial_\theta)
 +\big(\frac{1}{\textrm{sin}\theta}\partial_\phi
 -\frac{\textrm{cos}\theta}{\textrm{sin}\theta}\partial_\psi\big)^2\bigg]\varphi
 +\frac{4K^2}{r_\infty^2}\partial_\psi^2\varphi \bigg\}.
 \label{actionion}
\end{eqnarray}
Performing the partial wave decomposition
\begin{eqnarray}
 \varphi=\sum_{lm\lambda}\varphi_{lm\lambda}(\tau,
 \rho)e^{i\lambda\psi}e^{im\phi}S^{\lambda}_{lm}(\theta),
\end{eqnarray}
where $S(\theta)$ is the so called spin-weighted spherical function
satisfying the equation (recent paper can be found in
\cite{Berti2006prd})
\begin{eqnarray}
 \frac{1}{\textrm{sin}\theta}\partial_\theta(\textrm{sin}\theta\partial_\theta S)
 -\frac{(m-\lambda\cos\theta)^2}{\sin^2\theta}S
 +[l(l+1)-\lambda^2]S=0,
 \label{spin fuction}
\end{eqnarray}
The orthogonality condition reads
\begin{equation}
\int
S^{\lambda}_{lm}\overline{S}^{\lambda}_{l'm'}d\theta=\delta_{ll'}\delta_{mm'},
\end{equation}
where
\begin{equation}
\overline{S}^{\lambda}_{lm}=(-1)^{\lambda+m}S^{-\lambda}_{l(-m)}.
\end{equation}
Define the tortoise coordinate as
\begin{eqnarray}
 \frac{d\rho_*}{d\rho}=\frac{K}{F}.
\end{eqnarray}
With the tortoise coordinate $\rho_*$, we can write the action
(\ref{actionion}) as
\begin{eqnarray}
 S[\varphi]= &-&\frac{1}{2}\sum_{lm\lambda}\int d\tau d\rho_*
       K^{2}r_{\infty} \rho \varphi_{lm\lambda}^*
       \bigg[-(\partial_{\tau}-ieA_{\tau})^{2}\varphi_{lm\lambda}
    +\frac{1}{\rho^{2}K}\partial_{\rho_{*}}\rho^{2}K
      \partial_{\rho_{*}}\varphi_{lm\lambda}\nonumber\\
    &-&F[\frac{l(l+1)-\lambda^{2}}{\rho^{2}K^{2}}
       +\frac{4\lambda^{2}K^{2}}{r_{\infty}^{2}}]\varphi_{lm\lambda}\bigg]
\end{eqnarray}
where we have integrated the angle coordinate parts. Near the
horizon, the factor $F(\rho)$ vanishes. So, the last term can be
ignored. Back to the coordinate $\rho$, effective two-dimensional
action functional is simplified as
\begin{eqnarray}
 S[\varphi]= \frac{1}{2}\sum_{lm\lambda}\int d\tau
 d\rho Kr_{\infty}\rho^2\varphi_{lm\lambda}^*\bigg[-\frac{K}{F}
 (\partial_\tau-ieA_\tau)^2\varphi_{lm\lambda}+
 \partial_\rho \bigg(\frac{F}{K}\partial_\rho\varphi_{lm\lambda}\bigg)\bigg].
\end{eqnarray}
After undergoing the dimensional reduction near the horizon, each
partial wave of the scalar field can be effectively described by an
infinite collection of complex scalar field in the background of a
$(1+1)$-dimensional metric. The dilaton field $\Psi$, and the gauge
potential $A_\mu$:
\begin{eqnarray}
 &&ds^2=-\frac{F}{K}d\tau^2+\frac{K}{F}d\rho^2,\nonumber\\
 &&\Psi=K\rho^2,\;\;
 A_\tau=\pm\frac{\sqrt{3}}{2}\frac{\sqrt{\rho_+\rho_-}}{\rho},\;\;
 A_\rho=0.
\end{eqnarray}
For the $U(1)$ gauge current $J_\mu$, the consistent form of d = 2
abelian anomaly is considered in \cite{Robinson2005prl,Iso2006prl}
\begin{eqnarray}
 \nabla_\mu J^\mu =\pm
 \frac{e^2}{4\pi}\epsilon^{\mu\nu}\partial_\mu A_\nu,
 \label{current}
\end{eqnarray}
where $+(-)$ respectively corresponds to the left(right)-handed
fields and $\epsilon^{\mu\nu}$ is antisymmetric with
$\epsilon^{01}=1$. The current $J^\mu$ is not a covariant current.
However, a covariant current can be defined as \cite{Iso2006prl}
\begin{eqnarray}
 \tilde{J}^\mu =J^\mu \mp
 \frac{e^2}{4\pi}A_\lambda\epsilon^{\lambda\mu},
\end{eqnarray}
which satisfies
\begin{eqnarray}
 \nabla_\mu\tilde{J}^\mu=\pm\frac{e^2}{4\pi}\epsilon_{\mu\nu}F^{\mu\nu}.
\end{eqnarray}
The consistent current of (\ref{current}) can be written as
\begin{eqnarray}
 J^\mu=J^\mu_{(o)}\Theta_+(\rho)+J^\mu_{(H)}H(\rho),
\end{eqnarray}
where $\Theta_+(\rho)=\Theta(\rho-\rho_+-\epsilon)$ and
$H(\rho)=1-\Theta_+(\rho)$. $J^\mu_{(o)}(\rho)$, the current
outside the horizon, is conserved
\begin{eqnarray}
 \partial_\rho \big[J^\rho_{(o)}\big]=0.
\end{eqnarray}
While $J^\mu_{(H)}(\rho)$, the current near the horizon, satisfies
the anomalous equation
\begin{eqnarray}
 \partial_\rho \big[  J^\rho_{(H)}\big]=\frac{e^2}{4\pi}\partial_\rho A_\tau.
\end{eqnarray}
These two equations can be easily integrated as
\begin{eqnarray}
 J^\rho_{(o)}&=&c_{o},\nonumber\\
 J^\rho_{(H)}&=&c_{H}+\frac{e^2}{4\pi}(A_\tau(\rho)-A_\tau(\rho_+)).
\end{eqnarray}
where $c_o$ and $c_H$ are integration constants. The variation of
quantum effective action $W$ under a gauge transformation with gauge
parameter $\zeta$ is given by
\begin{eqnarray}
 -\delta W&=&\int d^2x \zeta\nabla_\mu J^\mu\nonumber\\
 &=&\int d^2x \zeta \bigg[\partial_\rho \big(\frac{e^2}{4\pi}A_\tau H(\rho)\big)
 +\delta(\rho-\rho_+-\epsilon)\big(
 (J^\rho_{(o)}-J^\rho_{(H)})+\frac{e^2}{4\pi}A_\tau\big)\bigg],
\end{eqnarray}
The coefficient of the delta function in the above equation should
vanish
\begin{eqnarray}
 c_o=c_H-\frac{e^2}{4\pi}A_\tau(\rho_+).
\end{eqnarray}
Imposing the condition that the covariant current
($\tilde{J}^{\rho}=J^{\rho}+\frac{e^{2}}{4\pi}A_\tau(\rho)H(\rho)$)
vanishes at the horizon, one can fix the value of the consistent
current at the horizon
\begin{eqnarray}
 c_H=-\frac{e^2}{4\pi}A_\tau(\rho_+).
\end{eqnarray}
The charge flux can be obtained
\begin{eqnarray}
 c_o=-\frac{e^2}{2\pi}A_\tau(\rho_+)=\frac{e^2}{2\pi}V,
\end{eqnarray}
which agrees with the current flow associated with the Hawking
thermal (blackbody) radiation including a chemical potential.
$V=-A_\tau(\rho_+)=\mp\frac{\sqrt{3}}{2}\sqrt{\frac{\rho_-}{\rho_+}}$
is the electrostatic potential of the black hole.

With the presence of gauge fields, the energy-momentum tensor does
not preserve the current conservation law and the corresponding
anomalous Ward identity is
$\nabla_{\mu}T^{\mu}_{\nu}=F_{\mu\nu}J^{\mu}+A_{\nu}$. Considering
the outside the horizon and near the horizon, the energy-momentum
tensor can be written as
\begin{eqnarray}
 T^\mu_\nu=T^\mu_{\nu(o)}\Theta_+(\rho)+T^\mu_{\nu(H)}H(\rho).
\end{eqnarray}
Outside the horizon, we have
\begin{eqnarray}
 \partial_\rho \big[T^\rho_{\tau(o)}\big]=J^\rho_{(o)}\partial_\rho
 A_\tau.
\end{eqnarray}
Near the horizon, we have the anomalous equation
\begin{eqnarray}
 \partial_\rho \big[ T^\rho_{\tau(H)}\big]=\big[J^\rho_{(H)}\partial_\rho
 A_\tau +A_\tau\partial_\rho J^\rho_{(H)}\big]+\partial_\rho
 N^\rho_\tau,
\end{eqnarray}
where $N^\rho_\tau$ is given by
\begin{eqnarray}
 N^\rho_\tau=\frac{1}{96\pi}\epsilon^{\mu\nu}\partial_\nu\Gamma^\rho_{\tau\mu}.
 \label{Nrhotau}
\end{eqnarray}
These two equations can also be integrated as
\begin{eqnarray}
 T^\rho_{\tau(o)}&=&a_o+c_o A_\tau,\nonumber\\
 T^\rho_{\tau(H)}&=&a_H+\int_{\rho_+}^\rho d\rho
 \partial_\rho\bigg(c_o
 A_\tau+\frac{e^2}{4\pi}A_\tau^2+N^\rho_\tau\bigg).
\end{eqnarray}
Under the infinitesimal general coordinate transformations, the
effective action changes as
\begin{eqnarray}
 -\delta W&=&\int d^2x \xi^\tau\nabla_\mu T^\mu_\tau\nonumber\\
 &=&\int d^2x \xi^\tau \bigg\{ c_o\partial_\rho A_\tau
 +\partial_\rho\bigg[\big(\frac{e^2}{4\pi}A_\tau^2+N^\rho_\tau\big)H\bigg]\nonumber\\
 &&+\delta(\rho-\rho_+-\epsilon)\bigg[(T^\rho_{\tau(o)}-T^\rho_{\tau(H)})
 +N^\rho_\tau+\frac{e^2}{4\pi}A_\tau^2\bigg]
 \bigg\}.
\end{eqnarray}
The coefficient of the delta function term should also vanish at the
horizon
\begin{eqnarray}
 a_o=a_H+\frac{e^2}{4\pi}A_\tau^2(\rho_+)-N^\rho_\tau(\rho_+).
\end{eqnarray}
We impose a vanishing condition for the covariant energy-momentum
tensor
($\tilde{T}^{\rho}_{\tau}=T^{\rho}_{\tau}
 +\frac{1}{192\pi}((\frac{F}{K})(\frac{F}{K})''-2((\frac{F}{K})')^{2})$)
at the horizon, which gives the equation
\begin{eqnarray}
 a_H=2N^\rho_\tau(\rho_+).
\end{eqnarray}
The total flux of the energy-momentum tensor is given by
\begin{eqnarray}
 a_o=\frac{e^2}{4\pi}A_\tau^2(\rho_+)+N^\rho_\tau(\rho_+),
\end{eqnarray}
From Eq. (\ref{Nrhotau}), we can calculate
\begin{eqnarray}
 N^\rho_\tau(\rho_+)=\frac{1}{192\pi}\big(\frac{F}{K}\big)'^2\big|_{\rho_+}
 =\frac{\kappa^2}{48\pi}.
\end{eqnarray}
So we have
\begin{eqnarray}
 a_o=\frac{e^2}{4\pi}V^2+\frac{\kappa^2}{48\pi},
\end{eqnarray}
where $V^2 = \frac{3}{4}\frac{\rho_{-}}{ \rho_{+}}$ and $\kappa$ is
the surface gravity.

\section{Quantum field in a Rotating Kaluza-Klein black hole with squashed horizon}
\label{totKK}

The five-dimensional rotating squashed Kaluza-Klein black hole is
described by\cite{Wang2006npb}
\begin{eqnarray}
ds^2=-dt^2+\frac{\Sigma_0}{\Delta_0}k(r)^2dr^2+\frac{r^2+a^2}{4}
[k(r)(\sigma^2_1+\sigma^2_2)+\sigma^2_3]+\frac{M}{r^2+a^2}(dt-\frac{a}{2}\sigma_3)^2,
\label{metric0}
\end{eqnarray}
with
\begin{eqnarray}
\sigma_1&=&-\sin{\psi} d\theta+\cos{\psi} \sin{\theta}d\phi,\nonumber\\
\sigma_2&=&\cos{\psi} d\theta+\sin{\psi} \sin{\theta}d\phi,\\
\sigma_3&=&d\psi+\cos{\theta}d\phi.\nonumber
\end{eqnarray}
The functions are given by
\begin{eqnarray}
\Sigma_0&=&r^2(r^2+a^2),\nonumber\\
\Delta_0&=&(r^2+a^2)^2-Mr^2,\\
k(r)&=&\frac{(r^2_{\infty}-r^2_+)(r^2_{\infty}-r^2_-)}{(r^2_{\infty}-r^2)^2},\nonumber
\end{eqnarray}
where $M$ and $a$ correspond to mass and angular momenta,
respectively. $r_+$ and $r_-$ are outer and inner event horizons of
the black hole. $r_\infty$ is the spatial infinity. The relations
between the parameters are $a^4=(r_+r_-)^2,\; M-2a^2=r^2_++r^2_-$.
It can be seen that the shape of black hole horizon is also deformed
by the parameter $k(r)$.

The determinant of the metric is
\begin{equation}
 g=-\frac{1}{64} k^{4}r^{2}(a^{2}+r^{2})^{2}\sin^{2}\theta,
\end{equation}
and non-zero metric coefficients read
\begin{eqnarray}
 &&g^{00}=-1-\frac{M(a^{2}+r^{2})}{\Delta_0},\nonumber\\
 &&g^{04}=g^{40}=-\frac{2aM}{\Delta_0},\nonumber\\
 &&g^{11}=\frac{\Delta_0}{k^{2}\Sigma_0},\nonumber\\
 &&g^{22}=\frac{4}{k(a^{2}+r^{2})}, \\
 &&g^{33}=\frac{4}{k(a^{2}+r^{2})\sin^{2}\theta},\nonumber\\
 &&g^{34}=g^{43}=-\frac{4\cos\theta}{k(a^{2}+r^{2})\sin^{2}\theta},\nonumber\\
 &&g^{44}=\frac{4(a^{2}-M+r^{2})}{\Delta_0}+\frac{4\cos^{2}\theta}{k(a^{2}+r^{2})\sin^{2}\theta}.\nonumber
\end{eqnarray}
We note that $g^{03}$ vanishes, which is different from that in
the usual five dimensional Kerr black hole. So, this implies that
there exists some special properties in such spacetime.

Now, we consider the complex scalar field in a rotating
Kaluza-Klein black hole background. The action functional is given
by
\begin{equation}
 S=\frac{1}{2} \int d^{5}x
 \sqrt{-g}\Phi^* \nabla^{2}\Phi+S_{int},\label{action}
\end{equation}
where $S_{int}$ includes mass and interaction terms. Near the
outer horizon, the first term of (\ref{action}) gives a dominant
contribution to the action and thus we can ignore the mass and
interaction terms $S_{int}$. And $\nabla^{2}$ is the
Laplace-Beltrami operator defined by
\begin{equation}
 \nabla^{2}\equiv
 \frac{1}{\sqrt{-g}}\partial_{\mu}(\sqrt{-g}g^{\mu\nu}\partial_{\nu}).
\end{equation}
Hence, near the outer horizon, the action becomes
\begin{eqnarray}
 S[\Phi]
  &=&\frac{1}{2}\int d^5x
         \Phi^{*}\bigg\{\sqrt{-g}g^{00}\partial_{t}^{2}
         +2 \sqrt{-g}g^{04}\partial_{t}\partial_{\psi}
         + \partial_{r} (\sqrt{-g}g^{11}\partial_{r}) \nonumber\\
   &&+\frac{kr \sin\theta}{2}\bigg[\frac{1}{\sin\theta}\partial_{\theta}(\sin\theta\partial_{\theta})
         +\big(\frac{1}{\textrm{sin}\theta}\partial_\phi
         -\frac{\textrm{cos}\theta}{\textrm{sin}\theta}\partial_\psi\big)^2 \\
   &&+\frac{k(a^{2}+r^{2})(a^{2}-M+r^{2})}{\Delta_0}\partial_{\psi}^{2}\bigg]\bigg\}\Phi. \nonumber
\end{eqnarray}
Performing the partial wave decomposition
\begin{equation}
 \Phi(t,r,\theta,\phi,\psi)
  =\sum_{\lambda lm}A^{\lambda}_{lm}(t,r)e^{im\phi
    +i\lambda\psi}S^{\lambda}_{lm}(\theta),
\end{equation}
where function $S(\theta)$ is the spin-weighted spherical function
and satisfies equation (\ref{spin fuction}). The physics near the
horizon can be effectively described by an infinite collection of
massless (1 + 1)-dimensional fields with the following action
\begin{eqnarray}
 S[\Phi]&=&\frac{1}{2}\sum_{\lambda lm}\int dt dr
           (A^{\lambda}_{lm})^{*}\bigg\{\frac{\sqrt{-g}}{\sin\theta}
             g^{00}\bigg(\partial_{t}
              +i\lambda\frac{2a M}{a^{4}+r^{4}+a^{2}(M+2 r^{2})}\bigg)^{2}\nonumber\\
         &&+\partial_{r}\bigg(\frac{\sqrt{-g}}{\sin\theta}g^{11}\partial_{r}\bigg)
           -\bigg[\frac{kr}{2}E_{\lambda lm}
             +\frac{\lambda^{2}k^{2} r (a^{2}+r^{2})^{2}}{2(a^{4}+r^{4}+a^{2}(M+2 r^{2}))}
             \bigg]\bigg\}A^{\lambda}_{lm}, \label{2dAction}
\end{eqnarray}
where $(A^{\lambda}_{lm})^{*}=2\pi(-1)^{l-m} A^{-\lambda}_{l (-m)}$
and $E_{\lambda lm}=-\{l(l+1)-\lambda^2\}$. Define the tortoise
coordinate as
\begin{eqnarray}
 \frac{dr_*}{dr}=\frac{1}{\Delta_{0}}.
\end{eqnarray}
The factor $\Delta_0=(r^2+a^2)^2-Mr^2$ vanishes near the horizon.
Carrying out the same method in Section (\ref{nonKK}), one will see
that the last term in the action (\ref{2dAction}) can be ignored.
Then the effective two-dimensional action (\ref{2dAction}) can be
simplified as
\begin{eqnarray}
 S[\Phi] =\frac{1}{2}\sum_{\lambda lm}\int dt dr
    \Psi (A^{\lambda}_{lm})^{*}
    \bigg\{g^{tt}  \big (\partial_{t}+ i\lambda A_{t}\big)^{2}
    +\partial_{r}\big(g^{rr}\partial_{r}\big) \bigg\} A^{\lambda}_{lm},
\end{eqnarray}
where
\begin{eqnarray}
 g^{tt}&=&-(g^{rr})^{-1}=
       -\frac{[(r^{2}+a^{2})^{2}+Ma^{2}]^{1/2}\Sigma_0^{1/2}k}{(r^2+a^2)^2-Mr^2} ,\\
 \Psi &=& \frac{1}{8} k r (a^{2}+r^{2})
      \frac{[(r^{2}+a^{2})^{2}+Ma^{2}]^{1/2}}{\Sigma_0^{1/2}}, \label{dilaton2} \\
 A_t &=& \frac{2a M}{(r^{2}+a^{2})^{2}+Ma^{2}}.\label{GaugePotential2}
\end{eqnarray}
Hence, after undergoing the dimensional reduction near the
horizon, each partial wave of the scalar field can be effectively
described by an infinite collection of complex scalar field in the
background of the dilaton field $\Psi$ (\ref{dilaton2}), the gauge
potential $A_\mu=(A_t,0)$ (\ref{GaugePotential2}), and a
$(1+1)$-dimensional metric described by
\begin{eqnarray}
 &&ds^2=-\frac{(r^2+a^2)^2-Mr^2}
              {[(r^{2}+a^{2})^{2}+Ma^{2}]^{1/2}\Sigma_0^{1/2}k}    dt^2
       +\frac{[(r^{2}+a^{2})^{2}+Ma^{2}]^{1/2}\Sigma_0^{1/2}k}
             {(r^2+a^2)^2-Mr^2}        dr^2.
\end{eqnarray}
Now, the total flux of the energy-momentum tensor for the rotating
Kaluza-Klein black hole with squashed horizon is calculated as
\begin{eqnarray}
 a_o&=&\frac{e^2}{4\pi}A_t^2(r_+)+N^\rho_t(r_+)\nonumber\\
    &=&\frac{e^2}{4\pi}\bigg(\frac{2a M}{a^{4}+r_+^{4}+a^{2}(M+2r_+^{2})}\bigg)^{2}
       +\frac{\pi}{12\beta^{2}},
\end{eqnarray}
where
\begin{eqnarray}
 \beta 
      &=&
      2\pi \frac{(r_{+}+r_{-})}{(r_{+}-r_{-})}
           \frac{(r_{\infty}^{2}-r_{-}^{2})}
                {(r_{\infty}^{2}-r_{+}^{2})}.
\end{eqnarray}
Note that the flux is proportional to $(T_{H})^{2}$ with
$T_{H}=1/\beta$ the Hawking temperature of the black hole. The
angular velocity on the event horizon is given by
\begin{eqnarray}
 \Omega=\frac{2a M}{a^{4}+r_+^{4}+a^{2}(M+2r_+^{2})}.
\end{eqnarray}
It is worth to point out that the Hawking temperature $T_H$ here is
different from the results in Refs. \cite{Wang2006npb} and
\cite{Chen2008prd}. In fact, this is because the different selection
of the scale of time. If considering a time coordinate
transformation $t=\sqrt{\alpha}\tau$ with
$\alpha=\sqrt{\frac{(r_{\infty}^{2}+a^{2})^{2}+Ma^{2}}{(r_{\infty}^{2}+a^{2})^{2}-Mr_{\infty}^{2}}}$,
one will obtain the same Hawking temperature with Ref.
\cite{Wang2006npb}:
\begin{eqnarray}
 T'_H=\alpha \frac{1}{\beta}=\frac{(r_{+}-r_{-})\sqrt{(r_{\infty}^{2}+r_{+}r_{-})^{2}+r_{+}r_{-}(r_{+}+r_{-})^{2}}}
          {  2\pi r_{+}  (r_{+}+r_{-})  (r_{\infty}^{2}-r_{-}^{2})}
    \sqrt{  \frac{r_{\infty}^{2}-r_{+}^{2}}
                 {r_{\infty}^{2}-r_{-}^{2}}  }.
\end{eqnarray}

\section{Summary}

Using the method of quantum anomalies, we obtained the Hawking flux
from (4+1)-dimensional Kaluza-Klein black hole with squashed
horizon. The result shows that there is no difference between the
squashed horizon and the normal one, this may be the reason that
they are equivalent in topology. The method adopted here only
considers the quantum anomalies near the horizon, so it is generally
believed to be universal. We also extended this method to the
(4+1)-dimensional rotating Kaluza-Klein black hole with squashed
horizon. By integrating the action given by a scalar field minimally
coupled to gravity in the background of black hole near the horizon,
we obtained a (1+1)-dimensional effect action. That means physics
near the horizon can be described with an infinite collection of
massless (1+1)-dimensional scalar fields. We also found that, after
the dimensional reduction, an effective $U(1)$ gauge field is
generated by an angular isometry. The azimuthal quantum number
$\lambda$ serves as the charge of each partial wave. The results
show that the gauge current and energy-momentum tensor fluxes are
exactly equivalent to Hawking radiation from the event horizon for
both the non-rotating and rotating Kaluza-Klein black holes with
squashed horizon.

\section*{Acknowledgement}

This work was supported by Program for New Century Excellent
Talents in University, the National Natural Science Foundation of
China(NSFC)(No. 10705013), the Doctoral Program Foundation of
Institutions of Higher Education of China (No. 20070730055), the
Key Project of Chinese Ministry of Education (No. 109153) and the
Fundamental Research Fund for Physics and Mathematics of Lanzhou
University (No. Lzu07002).


\begin{thebibliography}{99}


\bibitem{hawking}
 S. W. Hawking,
    Commun. Math. Phys. \textbf{43}, 199 (1975).

\bibitem{Parikh:1999mf}
 M. K. Parikh and F. Wilczek,
    Phys. Rev. Lett. {\bf 85}, 5042 (2000).


\bibitem{tunneling}
 E. C. Vagenas,
    Phys. Lett. B {\bf 503}, 399 (2001);
 A. J. M. Medved,
    Class. Quant. Grav.  {\bf 19}, 589 (2002);
 M. Angheben, M. Nadalini, L. Vanzo and S. Zerbini,
    JHEP {\bf 0505}, 014 (2005);
 J. Y. Zhang and Z. Zhao,
    JHEP {\bf 0510}, 055 (2005);
 Q. Q. Jiang and S. Q. Wu,
    Phys. Lett. B {\bf 635}, 151 (2006);
 M. Nadalini, L. Vanzo and S. Zerbini,
    J. Phys. A \textbf{39}, 6601 (2006);
 Q. Q. Jiang, S. Q. Wu and X. Cai,
    Phys. Rev. D {\bf 73}, 064003 (2006);
 S. Q. Wu and Q. Q. Jiang,
    {\em Hawking Radiation of Charged Particles as Tunneling from Higher Dimensional
       Reissner-Nordstrom-de Sitter Black Holes }, arXiv:hep-th/0603082;
 E. T. Akhmedov, V. Akhmedova and D. Singleton,
    Phys. Lett. \textbf{B 642}, 124 (2006);
 R. Li and J. R. Ren,
    Phys. Lett. B \textbf{661}, 370 (2008);
 R. Li, J. R. Ren and S. W. Wei,
    Class. Quant. Grav. \textbf{25}, 125016 (2008);
 D. Y. Chen, Q. Q. Jiang, X. T. Zu,
    Class. Quant. Grav. \textbf{25}, 205022 (2008);
 R. Li, J. R. Ren and D. F. Shi,
    Phys. Lett. B, \textbf{670}, 446 (2009);
 T. Zhu and J. R. Ren,
    Euro. Phys. J. C \textbf{62}, 413 (2009);
 T. Zhu, J. R. Ren and M. F. Li,
    JCAP \textbf{0908} 010 (2009);
 R. Banerjee and B. R. Majhi,
    Phys. Lett. B \textbf{662}, 62 (2008);
 R. Banerjee and B. R. Majhi
    JHEP \textbf{0806}, 095 (2008);
 B. R. Majhi,
   Phys. Rev. D 79, 044005 (2009);
 R. Banerjee, B. R. Majhi and D. Roy,
  {\em Corrections to Unruh effect in tunneling formalism and mapping with
    Hawking effect}, arXiv:0901.0466 [hep-th];
 B. R. Majhi and S. Samanta,
  {\em Hawking Radiation due to Photon and Gravitino Tunneling}, arXiv:0901.2258 [hep-th];
 R. Banerjee and S. Kulkarni,
  Phys. Rev. D \textbf{79}, 084035 (2009);
 E. T. Akhmedov, T. Pilling and D. Singleton,
 Int. J. Mod. Phys. D \textbf{17}, 2453 (2008);
 K. Srinivasan and T. Padmanabhan,
      Phys. Rev. D \textbf{60}, 024007 (1999);
 S. Shankaranarayanan,
     Phys. Rev. D \textbf{67}, 084026 (2003);
 R. Banerjee, B. R. Majhi and S. Samanta,
    Phys. Rev. D \textbf{77}, 124035 (2008);
 R. Banerjee and B. R. Majhi,
    Phys. Lett. B \textbf{674}, 218 (2009);
 S. K. Modak,
    Phys. Lett. B \textbf{671}, 167 (2009);
 S. Shankaranarayanan, T. Padmanabhan and K. Srinivasan,
    Class. Quant. Grav. \textbf{19}, 2671 (2002);
 R. Banerjee and B. R. Majhi,
    Phys. Lett. B \textbf{675}, 243 (2009);
 R. Banerjee, B. R. Majhi and E. C. Vagenas,
  {\em Quantum tunneling, energy-time uncertainty principle and black hole spectroscopy},
   arXiv:0907.4271[hep-th].


\bibitem{Robinson2005prl}
 S. P. Robinson and F. Wilczek,
    Phys. Rev. Lett.  {\bf 95}, 011303 (2005)

\bibitem{Iso2006prl}
  S. Iso, H. Umetsu and F. Wilczek,
    Phys. Rev. Lett.  {\bf 96}, 151302 (2006).

\bibitem{Vagenas2006jhep}
 E. C. Vagenas and S. Das,
    JHEP \textbf{0610}, 025 (2006);
 M. R. Setare,
    Eur. Phys. J. C \textbf{49}, 865 (2007);
 K. Xiao, W. B. Liu and H. B. Zhang,
    Phys. Lett. B \textbf{647}, 482 (2007);
 Z. B. Xu and B. Chen,
    Phys. Rev. D \textbf{75}, 024041 (2007);
 H. Shin and W. Kim,
    JHEP \textbf{0706}, 012 (2007);
 W. Kim and H. Shin,
    JHEP \textbf{0707}, 070 (2007);
 S. Das, S. P. Robinson and E. C. Vagenas,
    Int. J. Mod. Phys. D \textbf{17}, 533 (2008).

\bibitem{Jiang2007plb}
 Q. Q. Jiang and S. Q. Wu,
    Phys. Lett. B \textbf{647}, 200 (2007);
 Q. Q. Jiang, S. Q. Wu and X. Cai,
    Phys. Lett. B \textbf{651}, 58 (2007);
 Q. Q. Jiang, S. Q. Wu and X. Cai,
    Phys. Lett. B \textbf{651}, 65 (2007);
 Q. Q. Jiang, S. Q. Wu and X. Cai,
    Phys. Rev. D \textbf{75}, 064029 (2007);
 Q. Q. Jiang, S. Q. Wu and X. Cai,
    Phys. Rev. D \textbf{76}, 029904 (2007);
 Q. Q. Jiang,
    Class. Quant. Grav. \textbf{24}, 4391 (2007).

\bibitem{Murata2007prd}
 K. Murata and U. Miyamoto,
    Phys. Rev. D \textbf{76}, 084038 (2007).

\bibitem{Miyamoto2008prd}
 U. Miyamoto and K. Murata,
    Phys. Rev. D \textbf{77}, 024020 (2008);
 B. Chen and W. He,
    Class. Quant. Grav. \textbf{25}, 135011 (2008).

\bibitem{Ma0709.3684}
 Z. Z. Ma,
    Int. J. Mod. Phys. A \textbf{23}, 2783 (2008);
 C. G. Huang, J. R. Sun, X. N. Wu and H. Q. Zhang,
    Mod. Phys. Lett. A \textbf{23}, 2957 (2008);
 X. N. Wu, C. G. Huang and J. R. Sun,
    Phys. Rev. D \textbf{77}, 124023 (2008);


\bibitem{Wushuangqing2007cqg}
 S. Q. Wu and J. J. Peng,
     Class. Quant. Grav. \textbf{24}, 5123 (2007);
 J. J. Peng and S. Q. Wu,
     Chin. Phys. B \textbf{17}, 825 (2008).


\bibitem{Pen0709.0044}
 J. J. Peng, S. Q. Wu and Z. Y. Zhao,
    Class. Quant. Grav. \textbf{25}, 135001 (2008).

\bibitem{Wushuangqing0709.4074}
 S. Q. Wu  and Z. Y. Zhao,
 {\em Comment on "Hawking Radiation and Covariant Anomalies"},
 arXiv:0709.4074 [hep-th].

\bibitem{Pen0709.0157}
 J. J. Peng and S. Q. Wu,
    Gen. Rel. Grav. \textbf{40}, 2619 (2008);
 K. Lin, X. X. Zeng and S. Z. Yang,
    Chin. Phys. Lett. \textbf{25}, 390 (2008);
 Y. W. Han, X. X. Zeng and S. Z. Yang,
    Int. J. Theor. Phys. \textbf{47}, 2011 (2008);
 X. X. Zeng, S. Z. Yang and D. Y. Chen,
    Chin. Phys. B \textbf{17}, 1629 (2008);
 K. Lin, S. Z. Yang and X. X. Zeng,
    Chin. Phys. B \textbf{17},  2804 (2008);
 S. W. Chen, X. W. Liu, K. Lin, X. X. Zeng and S. Z. Yang,
    Commun. Theor. Phys. \textbf{50}, 537 (2008);
 X. W. Liu, X. X. Zeng, S. W. Chen, K. Lin and S. Z. Yang,
    Commun. Theor. Phys. \textbf{50}, 541 (2008).

\bibitem{Gangopadhyay2008prd}
 S. Gangopadhyay and S. Kulkarni,
    Phys. Rev. D \textbf{77}, 024038 (2008);
 S. Gangopadhyay,
    Phys. Rev. D \textbf{77}, 064027 (2008).

\bibitem{Kulkarni2008cqg}
 S. Kulkarni,
     Class. Quant. Grav. \textbf{25}, 225023 (2008).


\bibitem{Wushuangqing2008cqg}
 S. Q. Wu, J. J. Peng and Z. Y. Zhao,
     Class. Quant. Grav. \textbf{25}, 135001 (2008)

\bibitem{Kim2008}
 W. Kim, H. Shin and M. Yoon,
     J. Korean Phys. Soc. \textbf{53}, 1791 (2008).

\bibitem{Bonora2008jhep}
 L. Bonora and M. Cvitan,
    JHEP \textbf{0805}, 071 (2008);
 L. Bonora, M. Cvitan, S. Pallua and I. Smolic,
    JHEP \textbf{0812}, 021 (2008).


\bibitem{Shirasaka2008}
 A. Shirasaka and T. Hirata,
 {\em Higher Derivative Correction to the Hawking Flux via Trace
    Anomaly},
   arXiv:0804.1910 [hep-th];
 R. Becar, P. Gonzalez, G. Pulgar and J. Saavedra,
 {\em Hawking radiation via Anomaly and Tunneling method from Unruh's
     and Canonical acoustic black hole},
   arXiv:0808.1735 [gr-qc];
 A. P. Porfyriadis, Phys. Rev. D \textbf{79}, 084039, (2009);
 R. Banerjee and B. R. Majhi,
  Phys. Rev. D \textbf{79}, 064024, (2009);
 R. Banerjee,
 Int. J. Mod. Phys. D \textbf{17}, 2539 (2009);
 V. Akhmedova, T. Pilling, A. de Gill and D. Singleton,
  Phys. Lett. B \textbf{673}, 227 (2009).

\bibitem{Christensen1977prd}
S. M. Christensen and S. A. Fulling, Phys. Rev. D \textbf{15}, 2088
(1977).

\bibitem{Banerjee2008prd}
 R. Banerjee and S. Kulkarni,
    Phys. Rev. D \textbf{77}, 024018 (2008).

\bibitem{Banerjee2008plb}
 R. Banerjee and S. Kulkarni,
    Phys. Lett. B \textbf{659}, 827 (2008).


\bibitem{Papantonopoulos2008}
 E. Papantonopoulos and P. Skamagoulis,
  Phys. Rev. D \textbf{79},084022 (2009).

\bibitem{Isop2006prd}
 S. Iso, H. Umetsu and F. Wilczek,
    Phys. Rev. D \textbf{74}, 044017 (2006);
 S. Iso, T. Morita and H. Umetsu,
    JHEP \textbf{0704}, 068 (2007).

\bibitem{Ishihara2006ptp}
 H. Ishihara and K. Matsuno,
     Prog. Theor. Phys. \textbf{116}, 417 (2006);
 R.-G. Cai, L.-M., N. Ohta,
    Phys. Lett. B \textbf{ 639}, 354 (2006).

\bibitem{Wang2006npb}
 T. Wang,
    Nucl. Phys. B \textbf{756}, 86 (2006).

\bibitem{Chen2008prd}
 S. Chen, B. Wang and R. Su,
   Phys. Rev. D \textbf{77}, 024039 (2008).

\bibitem{Ishihara2007prd}
 H. Ishihara and J. Soda,
   Phys. Rev. D \textbf{76}, 064022 (2007).

\bibitem{Berti2006prd}
 E. Berti, V. Cardoso and M. Casals,
   Phys. Rev. D \textbf{73}, 024013 (2006).



\end{thebibliography}
\end{document}